\documentclass[12pt]{article}

\usepackage{epsfig}
\usepackage{amsmath}
\usepackage{color}
\newcommand{\tr}{^{\prime}}
\def\b#1{\mbox{\boldmath $#1$}}    

\newcommand{\diag}{{\rm diag}}      
\renewcommand{\th}{\theta}

\newcommand{\be}{\beta}
\newcommand{\de}{\delta}
\newcommand{\la}{\lambda}

\newcommand{\ga}{\gamma}

\usepackage{bm}
\usepackage{epsfig}

\begin{document}
\title{Assessment of school performance through a multilevel latent Markov Rasch model}
\author{Francesco Bartolucci\footnote{Department of Economics, Finance and
Statistics, University of Perugia, Via A. Pascoli, 20, 06123
Perugia}\hspace*{1mm}, Fulvia Pennoni\footnote{Department of
Statistics, University of Milano-Bicocca, Via Bicocca degli
Arcimboldi 8, 20126 Milano.}\hspace*{1mm} and Giorgio Vittadini
\footnote{Department of Quantitative Methods for Business and
Economic Sciences, University of Milano-Bicocca, Via Bicocca degli
Arcimboldi 8, 20126 Milano.}} \maketitle
\begin{abstract}
An extension of the latent Markov Rasch model is described for the
analysis of binary longitudinal data with covariates when subjects
are collected in clusters, e.g. students clustered in classes. For
each subject, the latent process is used to represent the
characteristic of interest (e.g. ability) conditional on the effect
of the cluster to which he/she belongs. The latter effect is modeled
by a discrete latent variable associated with each cluster. For the
maximum likelihood estimation of the model parameters we outline an
EM algorithm. We show how the proposed model may be used for
assessing the development of cognitive Math achievement. This
approach is applied to the analysis of a dataset collected in the
Lombardy Region (Italy) and based on test scores over three years of
middle-school students attending public and private schools.

\vspace*{0.2cm} \noindent {\sc Key words}: binary  longitudinal
data; EM algorithm;  empirical comparison; multilevel model; public
versus private schools; school effect.
\end{abstract}

\vspace*{0.5cm}

\section{Introduction}
Nowadays, many studies of the educational systems are focused on the
difference in scholastic achievement due to the presence of
particular teachers, schools, or educational conditions. For this
aim, growth linear models are typically used. These models associate
a trajectory to each student which is defined by a series of random
effects having a continuous distribution. The outcomes are evaluated
as average test scores or gain scores at the end of each school year
and are corrected on the basis of observed covariates. Among these
approaches, the one based on the hierarchical multilevel models also
known as random effects models (Snijders and Bosker, 1999;
Raudenbush and Bryk, 2002;  Dronkers and Robert, 2008) is able to
take into account the hierarchical structure of the data due to
students nested in classes and schools.

In the value added models (see the Spring 2004 issue of the {\em
Journal of Educational and Behavioural statistics} for a discussion
of VAM) student achievement is modeled as a linear additive function
of the full history of inputs received plus the student's innate
ability. These models consider the achievement level at the
beginning of the period as a covariate and the achievement level at
the end of the same period as an outcome. They have been extended in
the following directions: (\textit{i}) to account for the
non-compliance and missing values generated by failing to
participate in testing after the first year (Rubin {\em et al.},
2004; Lubienski and Lubienski, 2006, Lubienski  {\em et al}., 2008);
(\textit{ii}) to analyze the effects of a multiyear sequence of
instructional experiences (i.e. the reassignment of students to
teachers and classes at the beginning of each year) and the presence
of more than one potential outcome for each treatment;
(\textit{iii}) to use individual variables varying across time (Hong
and Raudenbush, 2008); (\textit{iv}) to obtain students' achievement
outcomes as latent variables underlying the observed achievement
scores in a single-year study (Goldstein {\em et al.}, 2007). In the
last proposal, the latent scores are the common `causes' of the
students' responses depurated by the effects of specific factors;
they are corrected on the basis of the influence of the covariates
and the multilevel structure of the data is taken into account.
However this proposal can be improved by considering that when the
latent scores which measure the achievement are determined by
factorial models, the true value added, due to a particular teacher
or school, cannot be completed determined. In fact, the  effects of
the latent scores on students' achievement is not net to their
ability  and to  the difficulty  of the items. Moreover the model
proposed by Goldstein {\em et al.} (2007) is not a longitudinal
model, as called for the value added approach.

With observational data, strong assumptions are necessary to
interpret the results in causal terms due to the non-random
mechanism of assignment, which gives rise to selection bias. For
this aim, different methods of analysis have been suggested (see
Rubin {\em et al}., 2004, and the references therein;  Mc Ewan,
2000; Naep, 2005, 2006 and Schneider {\em et al.}, 2007). However,
as shown in other studies (see Stuart, 2007, and Morgan and Winship,
2007), if the distribution of the available covariates in the sample
is very similar to that in the population, the results can sustain a
causal interpretation.

Motivated by an application based on a dataset collected in the
Lombardy Region of Italy, and concerning test scores on Mathematics
from standardized assessments over three years of middle school, we
propose a latent Markov (LM) model which attempts to study how
cognitive achievement changes over time depending on observable
covariates and the type of school attended. The proposed model is a
standard tool for the analysis of binary longitudinal data when the
interest is in describing individual changes with respect to a
certain latent status (for a review see Langeheine and van de Pol,
2002). In particular, we consider a version of the latent Markov
model in which the distribution of the response variable depends on
the corresponding latent variable as in the Rasch model (Rasch,
1961, Bartolucci {\em et al}., 2008b). Moreover, following the
formulation of Bartolucci and Lupparelli (2007) we extend the model
to take into account the multilevel structure of the data. We allow
the initial and the transition probabilities of the latent process
to depend on time-constant or time-varying covariates as in Vermunt
{\em et al.} (1999) and on a latent variable having the role of
capturing the heterogeneity between classes.

In principle, it would be possible to use a Rasch parameterization
for the ability  within a value added structure. However, we prefer
a multilevel LM model because the latent structure is more flexible
and the estimation may be carried out more easily. In fact, the
likelihood of the model may be computed by using a recursion taken
from the literature on hidden Markov models (MacDonald and Zucchini,
1997). On the basis of similar recursions, an
Expectation-Maximization (EM) algorithm (Baum {\em et al.}, 1970,
Dempster {\em et al}., 1977) may be implemented for the estimation
of the model parameters. This avoids the use of quadrature or Monte
Carlo methods. The proposed LM approach is also useful when it is
important to cluster subjects into a small number of groups
corresponding to different membership probabilities. To our
knowledge, a multilevel LM approach has not been previously proposed
to study the development of student achievement.

The remainder of the paper is organized as follows. To set the
context for our study, the next section gives some details on the
Italian Educational system and on the dataset used for the
application. Section 3 describes the multilevel extension of the LM
Rasch model with covariates and Section 4 describes its maximum
likelihood estimation and the related model selection strategy. In
Section 5 we show the results of the application of the proposed
approach to the dataset described in Section 2. In the concluding
section, we provide a final discussion of the main findings.
\section{Preliminaries}
In the Italian school system there are both public and private
schools serving the same functions. The Italian Constitution
expressly states that private schools must not impose burdens on the
State. Therefore, non-state schools receive funding from some local
and regional governments (with vouchers) and the national government
has declared its intention to promote equal treatment by the
legislation enacted in March 2000 (State. Law No. 62). With that
legislation the non-state schools may form part of the public
educational system and the private schools have been specified by a
new formula of `scuole paritarie' (private schools). In Italy, the
unitary character of the national educational system is protected
through the national definition of curriculum goals, timetables, and
specific learning objectives, but the curriculum implemented
nationally may be supplemented with elective courses.

In Lombardy, a higher percentage of pupils attends paritarie schools
than in any other Italian region. For example, in 2006 only 13\% of
student nationally were enrolled in paritarie schools compared to
22.6\% in Lombardy. Twenty-four percent of all Italian students
attending private schools are form Lombardy. Moreover, in Lombardy
in 2006, there were 177 private middle schools, with a total of 981
classes, and public middle schools numbered 1,038, with a total of
10,912 classes.

The schools in the regional sample we study in this paper were
selected by the Regional Research Institute on Education of Lombardy
(IRRE). The sample is taken from those schools of the region which
in 2003 participated the in Italian pilot study proposed by the
Ministry of Education and run by the Italian Institute for the
Evaluation of the Education System (INVALSI). That project was aimed
at detecting competencies on Reading (Italian language) and
Mathematics at the primary and secondary school-level. The schools
participated on voluntary basis. In the regional project promoted by
the IRRE, the schools were randomly selected among those belonging
to seven homogeneous metropolitan areas which do not include
particularly privileged and unprivileged inhabitants. These schools
were invited to administer the test to the same students in the same
classes for other two years at the end of each educational year. The
schools have also been invited to administer a questionnaire to the
cohort of the students in Grade 7 to evaluate their background
characteristics.

The sample we study is composed of a longitudinal cohort of 1,246
students who progressed from Grade 6 to Grade 8 during the three
study years. The students were 994 and 252 from, respectively, 13
public and 7 paritarie middle schools. The overall number of classes
is 77. It is important to stress that students are not placed into
classes based on their ability or achievement and that at the end of
Grade 8 students who have been admitted must pass the national
examination to obtain the licence which is necessary to attend the
high school.

A different sequence of dichotomously scored items was administered
at the end each educational year. As mentioned above, the test
scores for April 2003 (Grade 6) were taken from the INVALSI pilot
study. The test administrated in April 2004 (Grade 7) and May 2005
(Grade 8) were specifically  designed for each Grade within the
regional project. The questionnaires consisted of 28, 30 and 39
items, respectively. They included some items from out of Grade
level for vertical scaling. Among the items of the test for Grade 7,
seven items were replicated from the items of the test for Grade 6.
Among the items of the test for Grade 8, five items were replicated
from those of the test for Grade 7.

\section{The multilevel latent  Markov Rasch model}
In the following, we briefly review the LM Rasch model (Bartolucci {\em
et al.}, 2008b) and then we formulate its multilevel extension,
which has a structure suitable for the analysis of the dataset that
motivates the present paper.
\subsection{Latent Markov Rasch model}
The LM Rasch model may be seen as a version of the Wiggins's (1973)
LM model in which the distribution of the item responses, given the
ability, is based on a Rasch parametrization (Rasch, 1961). The main
advantage of this model, with respect to traditional IRT models, is
that it allows for transition of the subjects between the latent
classes associated with different levels of abilities, so as to take
into account the dynamics of the individual characteristics, which
typically arises in longitudinal studies.

Let $n$ denote the number of examinees, let $T$ denote the number of
time occasions and let $J_t$ denote the number of items administered
to the examinees at occasion $t$, with $t=1,\ldots,T$. For each
subject $i$, $i=1,\ldots,n$, the item responses are represented by
the random vector $\b Y_i^{(t)}$ having elements $Y_{ij}^{(t)}$,
$j=1,\ldots,J_t$. Also let $\b Y_i$ be the overall vector of
responses provided by this subject and suppose that individual
covariates, if available, are fixed and given. In this framework,
the basic assumptions of the LMR model may be summarized as follows:
\begin{itemize}
\item the vectors $\b Y_1,\ldots,\b Y_n$ of the responses provided by
the subjects in the sample are independent;
\item for each subject $i$, the response vectors $\b Y_i^{(t)}$, $t=1,\ldots,T$, are conditionally
independent given a latent process $V_i^{(1)},\ldots,V_i^{(T)}$
which follows a Markov chain with state space $\{1,\ldots,k\}$;
\item for each subject $i$ and occasion $t$, the random variables
$Y_{ij}^{(t)}$ are conditionally independent given $V_i^{(t)}$ and,
as in the Rasch model,
\begin{equation}
\la_j^{(t)}(v)=p(Y_{ij}^{(t)}=1|V_i^{(t)}=v) =
\frac{\exp(\th_v-\be_j^{(t)})}{1+\exp(\th_v-\be_j^{(t)})},\quad
v=1,\ldots,k,\label{eq:Rasch}
\end{equation}
where $\th_v$ is the ability level of the examinees in latent state
$v$ and $\be_j^{(t)}$ is the difficulty level of the item.
\end{itemize}

Note that the initial and transition probabilities of the latent
process, denoted respectively by $\pi_i(v)$ and $\pi_i(v_1|v_0)$,
can depend on the covariates through a logit or similar
parametrizations; see Vermunt {\em et al.} (1999). Moreover, it is
natural to include in the model the constraint
\begin{equation}
\th_1\leq\cdots\leq\th_k,\label{eq:con_increasing}
\end{equation}
so that the levels of each latent variable $V_i^{(t)}$ correspond to
increasing levels of ability and then the latent states have a
direct interpretation.
\subsection{Multilevel extension}
We now consider a multilevel structure in which the $n$ examinees
are collected in $H$ clusters that, in our application, correspond
to the classes in each school.
Every subject is then identified by
the pair of indices $hi$, with $h=1,\ldots,H$ and $i=1,\ldots,n_h$
and where $n_h$ is the dimension of cluster $h$. Accordingly, we
denote the vector of responses by $\b Y_{hi}^{(t)}$, when these are
referred to a specific occasion $t$, and by $\b Y_{hi}$ when
referred to the overall set of items. Each single element of these
vector is denoted by $Y_{hij}^{(t)}$ and we also denote by $\b Y_i$
the set of these random variables for $i=1,\ldots,n_h$,
$j=1,\ldots,J_t$ and $t=1,\ldots,T$.

In this framework, we propose a multilevel extension of the LM Rasch
model illustrated above. This extension closely recalls the
multilevel extension of the ordinary LM model proposed by Bartolucci
and Lupparelli (2007). This extension is based on the introduction
of the discrete latent variables $U_h$, with support
$\{1,\ldots,k_1\}$, which have the role of capturing the
heterogeneity between clusters in terms of their effect on the
ability level of each subject. In our application, the clusters
correspond to different classes of students and then the cluster
effect is due to different factors, such as teacher, number of
students, type of school; some of these covariates can be also
unobserved. The resulting model is based on the following
assumptions:
\begin{itemize}
\item the response vectors $\b Y_1, \ldots,\b Y_H$ are independent
(now $\b Y_h$ is referred to the responses for all subjects in
cluster $h$);
\item for each cluster $h$, the response vectors $\b Y_{hi}$, $i=1,\ldots,n_h$, are conditionally
independent given the latent variable $U_h$;
\item for each subject $hi$, the response vectors $\b Y_{hi}^{(t)}$, $t=1,\ldots,T$, are conditionally
independent given the latent process
$V_{hi}^{(1)},\ldots,V_{hi}^{(T)}$ which follows a Markov chain
state space $\{1,\ldots,k_2\}$;
\item for each subject $hi$ and occasion $t$, the response variables
$Y_{hij}^{(t)}$, $j=1,\ldots,J_t$, are conditionally independent
given $V_{hi}^{(t)}$ and their distribution is formulated as in
(\ref{eq:Rasch}), with the ability level $\th_1,\ldots,\th_{k_2}$
satisfying constraint (\ref{eq:con_increasing}).
\end{itemize}

\begin{figure}[h]\centering
\includegraphics[width=11cm]{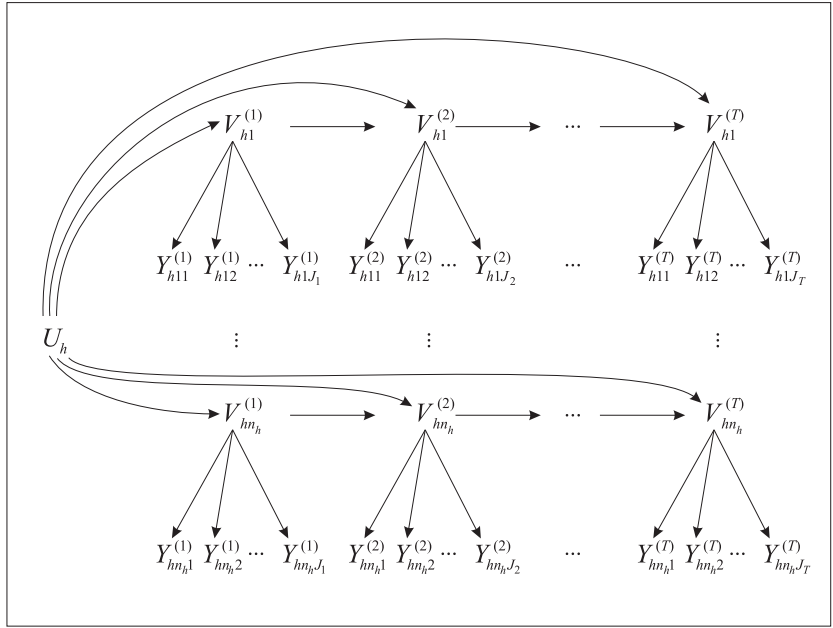}
\caption{\it Path diagram representing the multilevel LM
model.}\label{fig:path}
\end{figure}

The above assumptions lead to a dependence structure between the
latent and observable variables which is represented in the path
diagram depicted in  Figure \ref{fig:path} where, for simplicity,
covariates at individual and cluster levels are not indicated explicitly.

In order to complete the model specification, we need to formulate
the distribution of the latent variables given the available
covariates, which are assumed to be fixed and given. Those
covariates may be dummy, categorical or continuous. The covariates
referred the $h$-th cluster are collected in the vector $\b x_h$ and
the distribution of $U_h$ given these covariates is modeled through
the logit parametrization
\begin{equation}
\log\frac{\rho_h(u)}{\rho_h(1)}=\ga_{0u}+\b x_h\tr\b\ga_{1u},\quad
u=2,\ldots,k_1,\label{eq:logit_u}
\end{equation}
where $\rho_h(u)=p(U_h=u)$ and $\b\ga_{12},\ldots,\b\ga_{1k_1}$ are
vectors of regression coefficients of the same dimension as $\b x_h$
and $\ga_{02},\ldots,\ga_{0k_1}$ are the corresponding intercepts.

The covariates for subject $hi$ at occasion $t$ are denoted by $\b
z_{hi}^{(t)}$; these covariates are assumed to affect the initial
and the transition probabilities of the latent Markov process
$V_{hi}^{(1)},\ldots,V_{hi}^{(T)}$ by a parametrization based on
global logits. This type of parametrization is also adopted in a
similar context by Bartolucci {\em et al.} (2008a) and is motivated
by the ordinal nature of the variables $V_{hi}^{(t)}$. In
particular, for what concerns the initial probabilities
\[
\pi_{hi}(v|u) = p(V_{hi}^{(1)}=v|U_h=u)
\]
we assume
\begin{equation}
\log\frac{\pi_{hi}(v|u)+\cdots+\pi_{hi}(k_2|u)}{\pi_{hi}(1|u)+\cdots+\pi_{hi}(v-1|u)}=\de_{0u}+\de_{1v}+(\b
z_{hi}^{(1)})\tr\b\de_2,\quad u=1,\ldots,k_1,\quad
v=2,\ldots,k_2,\label{eq:logit_ini}
\end{equation}
where $\b\de_2$ is a vector of regression parameters of the same
dimension as each $\b z_{hi}^{(t)}$ which is common to every level
$v$; this is an usual assumption of models for ordinal variables
based on global logits (McCullagh, 1980). Moreover, the intercepts
$\de_{0u}$ depend on the level of $U_h$ and, in order to ensure
model identifiability, we let $\de_{01}\equiv 0$. On the other hand,
the intercepts $\de_{1v}$, depending on the level of $V_{hi}^{(1)}$,
must be in decreasing order, i.e.
$\de_{12}\leq\cdots\leq\de_{1k_2}$, to ensure the invertibility of
the global logit parametrization.

Finally, as regards the transition probabilities
\[
\pi_{hi}^{(t)}(v_1|u,v_0) =
p(V_{hi}^{(t)}=v_1|U_h=u,V_{hi}^{(t-1)}=v_0)
\]
we assume
\begin{equation}
\log\frac{\pi_{hi}^{(t)}(v_1|u,v_0)+\cdots+\pi_{hi}^{(t)}(k_2|u,v_0)}
{\pi_{hi}^{(t)}(1|u,v_0)+\cdots+\pi_{hi}^{(t)}(v_1-1|u,v_0)}=\eta_{0u}^{(t)}+\eta_{1v_0v_1}^{(t)}+(\b
z_{hi}^{(t)})\tr\b\eta_2^{(t)},\label{eq:logit_trans}
\end{equation}
with $u=1,\ldots,k_1$, $v_0=1,\ldots,k_2$, $v_1=2,\ldots,k_2$ and
$t=2,\ldots,T$. As above, $\b\eta_2^{(t)}$ is a common vector of
regression coefficients for the individual covariates, the
intercepts $\eta_{0u}$ depend on the level of $U_h$ (with
$\eta_{01}\equiv0$) and the intercepts $\eta_{0v_0v_1}$   depend on
the levels of $V_{hi}^{(t-1)}$ and $V_{hi}^{(t)}$ and must be
decreasing ordered in $v_1$ for each $v_0$.

Note that the covariates do not have a direct effect on the item
responses, but have a direct effect on the distribution the latent
variables $U_h$ and $V_{hi}^{(t)}$. As such, the support points
$\th_1,\ldots,\th_{k_2}$ are indeed interpretable as ability levels.
\subsection{Interpretation of the parameters}
A fundamental issue concerns how to interpret the model parameters.
First of all, the model assumes the existence of $k_2$ classes of
subjects which are ordered according to the ability level. The
ability level of class $v$ is denoted by $\th_v$. Moreover, for the
$j$-th item administered at the $t$-th occasion, $\be_j^{(t)}$ is
the difficulty level measured on the same scale of the ability.

Concerning the interpretation of the parameters for the distribution
of the latent variables, it is important to clarify that the
intercepts $\de_{1v}$ and $\eta_{1v_0v_1}$ in (\ref{eq:logit_ini})
and (\ref{eq:logit_trans}) are relatively less important. Of greater
interest are the parameters which characterize the clusters
according to their effect on the initial and transition
probabilities.
In this regard, the model assumes the existences of $k_1$ different
typologies of clusters. The effect of clusters of type $u$,
$u=1,\ldots,k_1$, on the initial probabilities is measured by
$\de_{0u}$ and the effect on the transition probability from
occasion $t-1$ to occasion $t$ is measured by $\eta_{0u}^{(t)}$.
This formulation allows us to consider time varying confounders as
well. Those clusters contribute the initial probabilities in
(\ref{eq:logit_ini}) and the transition probabilities
(\ref{eq:logit_trans}) in a way that it is possible to identify a
clear class effect which is time varying.

We can interpret similarly the regression coefficient in the vectors
$\b\de_2$ and $\b\eta_2^{(t)}$. With reference to our application,
for instance, if we find that $\de_{02}>\de_{01}$ and
$\eta_{02}^{(2)}<\eta_{01}^{(2)}$, this means that classes of type 2
have a better effect, with respect to classes of type 1, on the
ability of their students at the first occasion, but these classes
contribute less student's improvement from the first to the second
occasion. Moreover, suppose that for an individual covariate we have
a negative coefficient in $\b\de_2$, but a positive coefficient in
$\b\eta_2^{(2)}$. Then, as the value of the covariate increases, the
ability of the student at the first occasion decreases, but he/she
improves more consistently between the first and the second
occasion.

Finally, the parameter vectors $\b\ga_{1u}$ in (\ref{eq:logit_u})
are important for understanding how the distribution of the clusters
affects the $k_2$ different typologies described above. With
reference to our application, suppose that we have three typologies
of classes and that for a covariate describing some feature of these
classes we have a positive coefficient in both $\b\ga_{12}$ and
$\b\ga_{13}$.

This means that, as the value of the covariate increases, there is a
greater chance that the class is of type 2 (or of type 3) rather
than of type 1. The effect of the covariate on the probability that
the class is of type 3 rather than of type 2 depends on the
difference between the two regression coefficients. These effects
are not always easy to understand and in this case it may convenient
to directly consider the probability of each category of $U_h$ for
different levels of the covariate of interest. This is
straightforward when the covariate is a dummy for the class having a
particular attribute, such as being in a private rather than public
school.
\subsection{Computing the manifest distribution}
As in Bartolucci and Lupparelli (2007), the {\em manifest
distribution} of the response variables observed for each cluster
$h$ may be expressed as
\begin{equation}
p(\b Y_h=\b y_h) = \sum_u \la_h(u)\prod_i p(\b Y_{hi}=\b
y_{hi}|U_h=u),\label{eq:marg1}
\end{equation}
where $p(\b Y_{hi}=\b y_{hi}|U_h=u)$ may be efficiently computed by
a recursion which is known in the literature on hidden-Markov models
(Baum {\em et al.}, 1970, MacDonald and Zucchini, 1997). Details on
this recursion are given in Appendix 1.

Finally, the probability $p(\b Y_h=\b y_h)$ can be easily computed
through (\ref{eq:marg1}) and  the assumption of independence between
clusters implies that the manifest distribution of all the response
variables is given by
\[
p(\b Y_1=\b y_1,\ldots,\b Y_H=\b y_H)=\prod_hp(\b Y_h=\b y_h).
\]
\section{Likelihood inference}
The likelihood of the LM Rasch model may be expressed as
\[
\ell(\b\phi)=\sum_h \log p(\b Y_h=\b y_h),
\]
where $\b\phi$ is a short-hand notation for all model parameters
(see Section 3.2). In this section we show how this function may be
maximized, so as to obtain a maximum likelihood estimate of $\b\phi$
based on the observed sample, and we deal with related inferential
problems.
\subsection{Estimation}
Maximum likelihood estimation of the LM Rasch is carried out on the
basis of the EM algorithm (Baum {\em et al.}, 1970, Dempster {\em et
al.}, 1997). This algorithm is based on the {\em complete data}
likelihood, i.e. the likelihood that we would compute if we knew the
latent state of each subject at each occasion and the value of the
latent variable describing the effect of every cluster.

Let $w_h(u)$ be a dummy variable equal to 1 if cluster $h$ belongs
to latent class $u$, let $z_{hi}^{(t)}(v)$ be a dummy variable equal
to 1 if subject $i$ in cluster $h$ is in latent state $v$ at
occasion $t$ and let
$z_{hi}^{(t)}(v_0,v_1)=z_{hi}^{(t-1)}(v_0)z_{hi}^{(t)}(v_1)$ be a
dummy variable equal to 1 if subject $hi$ moves from state $v_0$ to
$v_1$ at occasion $t$. The complete data log-likelihood may be
expressed as
\begin{equation}
\ell^*(\b\phi) = \sum_h \sum_u w_h(u)\{\log [\rho_h(u)] +
m^*_h(u)\},\label{eq:comp_lik}
\end{equation}
where
\begin{eqnarray*}
m_h^*(u) &=&  \sum_i \sum_v z_{hi}^{(1)}(v)\log[\pi_{hi}^{(1)}(v|u)]+\\
&+&\sum_i\sum_{v_0}\sum_{v_1}\sum_{t>1}
z_{hi}^{(t)}(v_0,v_1)\log[\pi_{hi}^{(t)}(v_1|u,v_0)]+\\
&+&\sum_i\sum_v\sum_t
z_{hi}^{(t)}(v)[y_{hij}^{(t)}\log(\la_j^{(t)})+(1-y_{hij}^{(t)})\log(1-\la_j^{(t)})].
\end{eqnarray*}
Since the above dummy variables are not known, the EM algorithm
alternates the following two steps until convergence:
\begin{itemize}
\item {\bf E-step}: compute the conditional expected
value of the dummy variables $w_h(u)$, $z_{hi}^{(t)}(v)$ and
$z_{hi}^{(t)}(v_0,v_1)$ given the observed data and the current
value of the parameters;
\item {\bf M-step}: maximize the conditional expected value of
$\ell^*(\b\phi)$ obtained by substituting each dummy variable in
(\ref{eq:comp_lik}) with the corresponding expected value obtained
from the E-step; the resulting log-likelihood is denoted by
$\tilde{\ell}^*(\b\phi)$.
\end{itemize}

The conditional expected value of $w_h(u)$ corresponds to the
posterior probability $\tilde{w}_h(u)=p(U_h=u|\b Y_h=\b y_h)$ and
then, at the E-step, it is computed as
\[
\tilde{w}_h(u)=\la_h(u)\prod_i p(\b Y_{hi}=\b y_{hi}|U_h=u)/p(\b
Y_h=\b y_h).
\]
Similarly, we have $\tilde{z}_{hi}^{(t)}(v)=p(V_{hi}^{(t)}=v|\b
Y_h=\b y_h)$ and
$\tilde{z}_{hi}(v_0,v_1)=p(V_{hi}^{(t-1)}=v_0,V_{hi}^{(t)}=v_1|\b
Y_h=\b y_h)$ which are computed as
\begin{eqnarray*}
\tilde{z}_{hi}^{(t)}(v)&=&\sum_u
\tilde{z}_{hi}^{(t)}(v|u)\tilde{w}_h(u),\\
\tilde{z}_{hi}^{(t)}(v_0,v_1)&=&\sum_u\tilde{z}_{hi}^{(t)}(v_0,v_1|u)
\tilde{w}_h(u),
\end{eqnarray*}
where the conditional probabilities
\begin{eqnarray*}
\tilde{z}_{hi}^{(t)}(v|u)&=&p(V_{hi}^{(t)}=v|U_h=u,\b Y_h=\b y_h)\\
\tilde{z}_{hi}^{(t)}(v_0,v_1|u)&=&p(V_{hi}^{(t-1)}=v_0,V_{hi}^{(t)}=v_1|U_h=u,\b
Y_h=\b y_h)
\end{eqnarray*}
may be obtained by certain recursions which are illustrated in
Appendix 2.

Finally, the M-step is based on standard iterative algorithms to
maximize each component of $\tilde{\ell}^*(\b\phi)$. These
algorithms are the same as those used to estimate a multinomial
logit model on the basis of a weighted log-likelihood.

It is important to mention that, as typically happens for latent
variable models, the likelihood of the proposed model may be
multimodal and has a number of local maxima which increases with the
number of latent variables and that of the states. It is thus
crucial to choose the initial values of the EM algorithm
appropriately. In particular, we select the intercepts corresponding
to the different levels of the latent variables $U_h$ and
$V_{hi}^{(t)}$ on a grid of, respectively, $k_1$ and $k_2$
equispaced points around 0. Moreover, all the regression
coefficients for the covariates are fixed at 0, whereas the
difficulty levels of the items are chosen on the basis of the
observed frequencies of correct responses.
\subsection{Model selection and hypothesis testing}

For model selection, we rely on the Bayesian Information Criterion
(BIC; Schwarz, 1978), which is based on the index
\[
BIC = -2 \ell(\hat{\b\phi}) + r \log(n),
\]
where $\ell(\hat{\b \phi})$ is the maximum log-likelihood of the
model of interest and $r$ is the number of parameters;
the latter obviously depends on both $k_1$
and $k_2$. According to this criterion, the optimal combination of $k_1$ and $k_2$
is the one corresponding to the model with the smallest BIC value.

For testing a hypothesis on the parameters, we rely on the
likelihood ratio statistic
$D=-2[\ell(\hat{\b\phi}_0)-\ell(\hat{\b\phi})]$, where
$\hat{\b\phi}_0$ is the estimate of the parameter vector under the
hypothesis of interest, which can be computed through the same EM
algorithm described in section 4.1. To compute the standard errors
for the parameter estimates we rely on a method similar to the
likelihood profiling method (Meeker and Escobar, 1995). In
particular, for the estimate $\hat{\phi}_h$ of the parameter
$\phi_h$, we first compute the likelihood ratio statistic $D_h$ for
testing the hypothesis $H_0:\phi_h=0$ and then we compute the
standard error $se(\hat{\phi}_h)$ as $|\hat{\phi}_h|/\sqrt{D_h}$. In
this way, the conclusion of the Wald test for $H_0$ based on the
statistic $\hat{\phi}_h/se(\hat{\phi}_h)$ is guaranteed to be the
same as the test based on the statistic $D_h$.
\section{Application to the Lombardy dataset}
We now illustrate the analysis of the  longitudinal patterns of
achievement levels in Mathematics measured by the tests
administrated at the end of each school year between the two
subgroups of students attending public and paritarie schools.

Table \ref{Table1} presents the frequency distribution of the
available characteristics  of the public and paritarie middle
schools in 2003 at population level on the selected areas of
Lombardy. Table \ref{Table2} shows the corresponding sample
distributions, including a dummy variable related to the years since
school opened. It can be seen that the sample and the population
distributions look similar for both types of school. In both cases,
the public schools enroll more students and have a higher
student-teacher ratio. Table \ref{Table3} concerns the social
background characteristics of the students. It reports the
percentage values of father and mother level of education and the
percentage of missing responses.
\subsection{Model fitting and Results}
We here report the results obtained by applying the multilevel Rasch
LM model to the available dataset. We fitted the proposed model
including the students and the school covariates. We also included
two dummy variables to account for the student missing responses on
the questions related to father and mother levels of education.

We fitted the model for a different number of latent states at
cluster level ($k_1$), ranging form 1 to 5, and individual level
($k_2$), ranging from 1 to 7. Table  \ref{table4} shows the results
of the fitted models reporting the maximum log-likelihood
$(\hat{\ell}_{k_1,k_2})$ of the estimated model,  the value attained
by the BIC index and the number of parameters. We observe that the
lowest value of the BIC index corresponds to four typologies of
clusters ($k_1=4$) and six math ability levels ($k_1=6$). We then
identify six subgroups of students with different levels of ability
and four different types of school classes.

\begin{figure}[h]\centering
\includegraphics[width=12cm,clip=false]{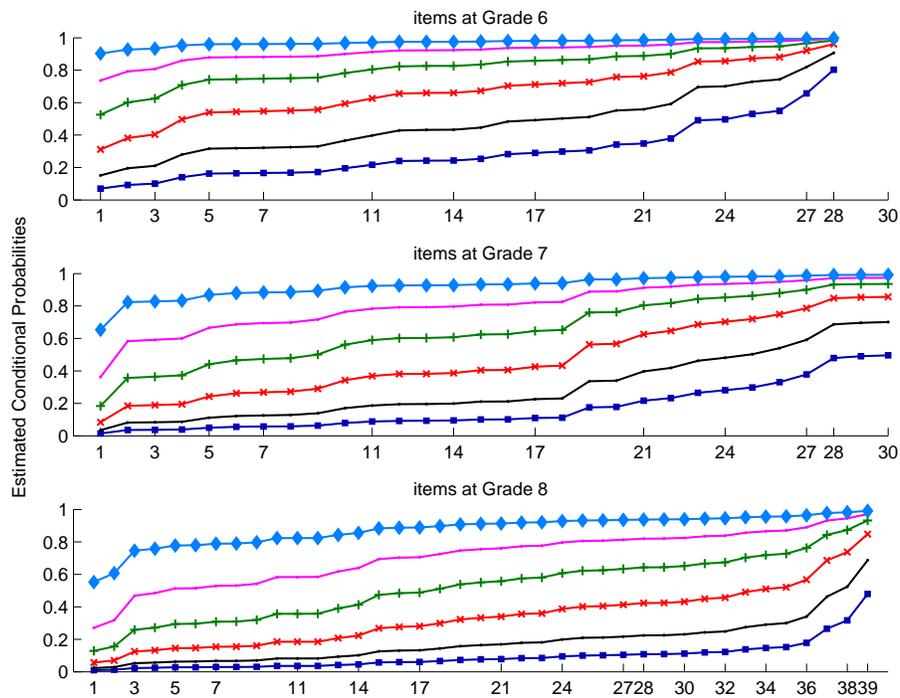}
\caption{\em Ordered estimated conditional probabilities for latent
classes 1-5: \textbf{blue square}, class 1;  \textbf{black bullet}
class 2; \textbf{red times}, class 3; \textbf{green plus}, class 4;
\textbf{pink bullet}, class 5; \textbf{heavenly diamond}, class 6.
The first graph form the top is referred to the items administrated
at Grade 6, the second at Grade 7 and the third at Grade
8.}\label{Figure2}
\end{figure}

The estimated abilities for each latent state are reported in Table
\ref{table5}. These abilities range from the lowest to the highest
levels; note that the ability of the first class is constrained to
be 0 to guarantee identifiability. These results  are in accordance
with the six student proficiency levels in Mathematics identified in
the OCSE-PISA reports (see for example OECD, 2007). They may
represent some specific types of task in math that a student is
likely to perform successfully. A better interpretation of these
latent classes can be gained by looking at the estimated conditional
probabilities parameterized  through a logit function of the
abilities and of the item difficulties. They are depicted  in Figure
\ref{Figure2} for each level of ability and according to the
different item which has been administrated at each grade.
From this figure we can read the probability of responding correctly
to each set of items for each grade for a student belonging to one
of the six latent classes. They are ordered from the lowest to the
highest in each graph.

As we would expect, these probabilities are higher for the items
administrated at Grade 6 (top-most graph) and are lower for the
items administrated at Grade 8 (bottom graph). It means that the
difficulty of the items is increasing over time. The items which
have been replicated over time share the same value of those
probabilities. For example item number 2 on the top graph is
replicated at Grade 7 and it corresponds to item number 13 of middle
graph. Additional observations can be drawn from this figure. For
example the probability of responding correctly to the items
administered at Grade 6 ranges between 0.8 and 1 for the students
with the highest ability. It ranges, instead, between 0 and 0.8 for
the students with the lowest  ability level. This means that they
are specially tailored to measures the abilities of the less capable
students.

Table \ref{table6} displays the estimates of the intercepts and the
regression coefficients for the logistic model at cluster level,
which is based on parametrization (\ref{eq:logit_u}). There are
three ordered intercepts, one for each of the three clusters
identified by the letters {\em B}, {\em C} and {\em D}. The equality
restrictions to zero have been imposed on the parameters of the
first latent class ({\em A}) to make the model identifiable. The
other estimated parameters are the regression coefficients for the
covariate type of school labeled with 1, to the ratio between
students and teachers labeled with 2, and to the dummy variable
indicating the years of activity of the school labeled with 3. As
the value of the covariate ratio between student and teachers
increases, there is less chance that the class is of type {\em A}
rather than of type {\em B}. On the other hand, as the value of the
years of activity of the school increases there is more chance that
the class is of type {\em A} rather than of type {\em B}.

To better interpret these estimates, it is convenient to consider
the probability of each cluster for different levels of the
covariate of interest. For example the average class probability of
belonging  to each cluster  is reported in Table \ref{table7}. It
indicates that 78\% of the classes of the paritarie schools belong
to cluster {\em A} and 23\% to cluster {\em D}, whereas the
percentage is 32\% and 6\% for the classes of the public middle
schools. On the basis on these results we conclude that the classes
of type {\em A} are prevalently those of paritarie schools with
small values of the ratio between students and teachers and with
years of activity higher than eighteen. The classes of type {\em B}
are mainly in public schools with different years of activity and
values of the ratio between students and teachers higher than eight.
The classes of type {\em C} are mainly
in public schools with different values of years of activity and of
the ratio between students and teachers. The classes of type {\em D}
are mainly in paritarie schools with years of activity less than
eighteen and with low values of the ratio between students and
teachers.

Table \ref{table8} displays the estimates of the intercepts and the
regression coefficients for the initial probabilities of the latent
Markov process. These parameter estimates can be interpreted on the
basis of formula (\ref{eq:logit_ini}).  In particular, there are
three ordered intercepts corresponding to the effect on the initial
probability of the clusters  {\em B}, {\em C} and {\em D}. Therefore
the classes of type {\em B} help less to increase the math ability
on the first year of the middle school compared to the classes of
the other clusters. The classes helping more on the first year are
those of type {\em C}.

Table \ref{table9}  shows the effects of the same variables on the
transition probabilities, from Grade 6 to 7 and Grade 7 to 8, of the
latent Markov process. On the basis of the estimated coefficients we
can state that the classes of type {\em B} contribute less on the
math ability of their students than those of type {\em C} and the
classes of type {\em  D} have a high positive effect on the math
ability from Grade 6 to Grade 7. However if we consider the
estimated coefficients related to the transition form Grade 7 to 8,
type {\em C} classes contribute the most to students' math ability.

Looking at the estimated regression coefficients related to the
covariate level of  education of the father we can see that the
ability of the students increases for those having higher educated
fathers. The magnitude of this increase is less strong than
on Grade 6 and is quite the same for the transition form Grade 6 to
7 and from Grade 7 to 8. For the global logit on the transition
probablities from Grade 7 to 8 the variables related to the missing
values are significant as well. On the basis of the estimates in
Table \ref{table9} we conclude that the contribution of the level of
education of the father is always inferior of the contribution of
type of class on the math achievement.

Finally, we provide a comparison of the above results with some
descriptive statistics on the student's scores of the sample across
grades. In particular, Table \ref{Table10} and Table \ref{Table11}
report, for public and paritarie middle schools, the empirical
transition matrices obtained by dividing the sum of the scores for
each subject in each grade into six classes of score. Looking at
these probabilities from Grade 6 to Grade 7 and from Grade 7 to
Grade 8 for both types of schools it can be noticed that the
students do not improve their abilities as much as we might expect.
There is not a transition towards state characterizing higher scores
but there is great persistence in the same state. This is also true
for the transition from Grade 7 to Grade 8 for both types of
schools. For both the paritarie and public schools, students with
less ability have some chance of becoming better performers at the
end of the middle school. Individuals attending public schools who
are in the first knowledge state show a probability of 0.67 of
moving to a better knowledge state form Grade 6 to  Grade 7 and a
probability of 0.35 of moving from state 2 at Grade 7 to state 3 at
Grade 8. In the paritarie schools there are more students with the
highest ability level at Grade 8 compared to the public schools. The
empirical transition matrices seem to have a tridiagonal structure:
the transition is possible only below or above the diagonal.

\section{Conclusions}
We propose a multilevel extension of the LM Rasch model for the
analysis of longitudinal data derived from the repeated
administration of binary test items to students attending public and
private middle schools in the Lombardy Region of Italy. The items
are aimed at assessing math knowledge of the students during the
three years of middle school. Taking into account that student
actual knowledge and the potential to increase such knowledge
depends on prior knowledge and socio-cognitive aspects, such as
family and school, we propose an alternative method to growth
models.

We show how the multilevel extension of the LM Rasch model allows us
to make a comparison between two types of schools with different
pupil achievement. The model assumes the existence of a latent
Markov chain for the ability level dynamics and it allows us to
model the probability of individual changes over time, while taking
into account the hierarchical structure of the data. It allows us to
flexibly parameterize the conditional distribution of the vector of
the response variables in order to take into account the different
number of items administered at each grade and the fact that items
may be replicated at different occasions.

We have shown that the acquisition of mathematical knowledge is a
result of the differences between student's background and personal
behavior. Moreover the rate of change that brings the student from
one knowledge level to the following
one may also depend on the quality of the school. When the school
tends to its task, family background is less influential on student
results. Therefore we could also conclude that the lower the
parental education is, the more the school helps.

Our results demonstrate that the model on which our approach is
based can describe the main relationships in the data with a rather
parsimonious structure. Moreover it takes into account that
cognitive achievement changes over time in relation to the
background variables of the student and the class type.

A causal interpretation can be given to the estimated regression
coefficients as the school covariates appear to have the same
distribution on the population of the school in Lombardy. Obviously,
in drawing conclusions on the basis of the application we have to
consider that schools participate on a voluntary basis in the IRRE
project from which the available data have been collected. This
could have determined a selection bias. For instance, it is possible
that the choice to participate in the study was only made by the
best organized schools with the most qualified teachers.

\section*{Appendices}

\subsection*{Appendix 1: computing manifest probabilities}
Following Bartolucci (2006) and Bartolucci {\em et al.} (2007), we
describe the recursion to compute $p(\b Y_{hi}=\b y_{hi}|U_h=u)$ by
using the matrix notation. This makes its implementation easier in
most mathematical and statistical packages.

Let $\b p_{hi}^{(t)}$ be a column vector with elements
\begin{eqnarray*}
p(\b Y_{hi}^{(t)}=\b y_{hi}^{(t)}|V_{hi}^{(t)}=v)&=&\prod_j
p(Y_{hij}^{(t)}=y_{hij}^{(t)}|V_{hi}^{(t)}=v)=\\
&=&\prod_j
[\la_j^{(t)}]^{y_{hij}^{(t)}}[1-\la_j^{(t)}]^{1-y_{hij}^{(t)}},\quad
v=1,\ldots,k_2,
\end{eqnarray*}
and consider the vector $\b q_{hi}^{(t)}(u)$ with elements
\[
p(\b Y_{hi}^{(1)}=\b y_{hi}^{(1)},\ldots,\b Y_{hi}^{(t)}=\b
y_{hi}^{(t)}|U_h=u,V_{hi}^{(t)}=v),\quad v=1,\ldots,k_2.
\]
The recursion mentioned above allows us to compute this vector as
\[
\b q_{it}(u) = \left\{\begin{array}{ll}
\textrm{diag}(\b p_{hi}^{(1)}) \b \pi_{hi}(u) & \mbox{ for } t=1,\\
\textrm{diag}(\b p_{hi}^{(t)}) [\b\Pi_{hi}^{(t)}(u)]\tr \b q_{i,t-1}(u) & \mbox{ for } t=2,\ldots,T,\\
\end{array}\right.
\]
where the vector $\b\pi_{hi}(u)$ has elements $\pi_{hi}(v|u)$,
$v=1,\ldots,k_2$, and the matrix $\b\Pi_{hi}^{(t)}(u)$ has elements
$\pi_{hi}^{(t)}(v_1|u,v_0)$, $v_0,v_1=1,\ldots,k_2$. At the end of
the recursion, we obtain $\b q_{iT}(u)$; the sum of the elements of
this vector is equal to $p(\b Y_{hi}=\b y_{hi}|U_h=u)$.

\subsection*{Appendix 2: computing posterior probabilities}
Let $\b z^{(t)}_{hi}(u)$ be the column vector with elements
$\tilde{z}_{hi}^{(t)}(v|u)$, $v=1,\ldots,k_2$, and and $\b
Z^{(t)}_{hi}(u)$ be the matrix with elements
$\tilde{z}_{hi}^{(t)}(v_0,v_1|u)$, $v_0,v_1=1,\ldots,k_2$. These may
be computed through the following backward recursion (see Levinson
\textit{et al.}, 1983, and MacDonald and Zucchini, 1997, Bartolucci
{\em et al.}, 2007):
\begin{eqnarray*}
\b z^{(t)}_{hi}(u) &=& \diag[\b q_{hi}^{(t)}(u)]\b
r_{hi}^{(t)}/p(\b Y_{hi}=\b y_{hi}|U_h=u),\\
\b Z^{(t)}_{hi}(u) &=& \diag[\b
q_{hi}^{(t-1)}(u)]\b\Pi_{hi}^{(t)}(u)\diag(\b
p_{hi}^{(t)})\textrm{diag}(\b r_{hi}^{(t)})/p(\b Y_{hi}=\b
y_{hi}|U_h=u),
\end{eqnarray*}
where
\[
\b r_{hi}^{(t)}(u) = \left\{\begin{array}{ll}
\b 1_{k_2}& \mbox{ if } t=T,\\
\b\Pi^{(t+1)}\textrm{diag}[\b p_{hi}^{(t+1)}(u)]\b r_{hi}^{(t+1)}(u)
& \mbox{ otherwise},
\end{array}\right.
\]
with $\b 1_{k_2}$ denoting a column vector of $k_2$ ones.
\section*{References}
{\small
\begin{description}
\item Bartolucci, F. (2006). Likelihood inference for a class of
latent Markov models under linear hypotheses on the transition
probabilities. {\em Journal of the Royal Statistical Society, series
B}, {\bf 68}, 155--178.\vspace*{-0.25cm}
\item Bartolucci, F. and Lupparelli, M. (2007). The multilevel latent Markov
model. Proceedings of the International Workshop on Statistical
Modelling, Barcelona, July 2007, pp. 93-98.\vspace*{-0.25cm}
\item Bartolucci, F., Lupparelli, M. and Montanari, G. E. (2008a),
Latent Markov model for binary longitudinal data: an application to
the performance evaluation of nursing homes, {\em to appear in
Annals of Applied Statistics}, available at
\verb"www.stat.unipg.it\bartolucci\".\vspace*{-0.25cm}
\item Bartolucci, F. and Pennoni, F. (2007). A class of latent Markov models for
capture-recapture data allowing for time, heterogeneity, and
behavior effects. {\em Biometrics}, {\bf 63}, 568--568.
\vspace*{-0.25cm}
\item Bartolucci, F., Pennoni, F. and
Francis, B. (2007). A latent Markov model for detecting patterns of
criminal activity. {\em Journal of the Royal Statistical Society,
series A}, {\bf 170}, 115--132.\vspace*{-0.25cm}
\item Bartolucci,  F.,  Pennoni,  F., Lupparelli, M. (2008b).
Likelihood inference for the latent Markov Rasch model. In: C.
Huber, N. Limnios, M. Mesbah, M. Nikulin (Eds.), {\em Mathematical
Methods for Survival Analysis, Reliability and Quality of Life},
239--254, Wiley.\vspace*{-0.25cm}
\item Baum, L. E., Petrie, T., Soules, G. and Weiss, N. (1970), A
maximization technique occurring in the statistical analysis of
probabilistic functions of Markov chains, {\em Annals of
Mathematical Statistics}, {\bf 41}, pp. 164-171.\vspace*{-0.2cm}
\item Dronkers J. and  Robert P. (2008).  Differences in Scholastic
Achievement of Public, Private Government-Dependent, and Private
Independent Schools.  {\em Educational Policy}  {\bf 22},
541-577.\vspace*{-0.25cm} %
\item Goldstein H., Bonnet G. and Rocher T. (2007). Multilevel
Structural Equation Models for the Analysis of Comparative Data on
Educational Performance. {\em Journal of Educational and Behavioral
Statistics},  {\bf 32}, 252-286. \vspace*{-0.25cm}
\item Dempster, A. P., Laird, N. M. and Rubin, D. B. (1977).
Maximum likelihood from incomplete data via the EM algorithm (with
discussion). {\it J. R. Statist. Soc. series B}, {\bf 39}, 1-38.
\vspace*{-0.25cm} %
\item Hong G. and Raudenbush S. W. (2008). Causal Inference for
Time-Varying Instructional Treatments. {\em Journal of Educational
and Behavioral Statistics}, {\bf 33}, 333-362.\vspace*{-0.25cm}
%
%
%
\item Langeheine, R. and van de Pol, F. (2002), Latent Markov Chains, in {\em
Advances in Latent Class Analysis}, A. L. McCutcheon and J. A.
Hagenaars editors, University Press, Cambridge.\vspace*{-0.2cm}
%
%
\item Levinson, S. E., Rabiner, L. R. and Sondhi, M. M. (1983), An
introduction to the application of the theory of probabilistic
functions of a Markov process to automatic speech recognition, {\em
Bell System Technical Journal}, {\bf 62}, pp.
1035-1074.\vspace*{-0.2cm}
%
%
%
\item Lubienski, S.T. and
Lubienski, C. (2006). School sector and academic achievement: A
multi-level analysis of NAEP mathematics data. {\em American
Educational Research Journal}, {\bf 43}, 651-698.\vspace*{-0.25cm}
\item Lubienski, C., Crane, C. C. and Lubienski, S. T. (2008). What Do
We Know About School Effectiveness?  Academic Gains in a Value-Added
Analysis of Public and Private Schools.  {\em Phi Delta Kappan},
{\bf 89}, 689-695.\vspace*{-0.25cm}
\item McCullagh, P. (1980). Regression models for ordinal data (with discussion).
{\em Journal of the Royal Statistical Society, series B}, {\bf 42},
109-142.\vspace*{-0.25cm}
\item MacDonald, I. and Zucchini, W. (1997). {\em Hidden Markov
and Other Models for Discrete-valued Time Series}. London: Chapman
\& Hall.\vspace*{-0.25cm}
\item Meeker, W. Q. and Escobar L. A.  (1995). Teaching about approximate confidence regions based on maximum likelihood estimation.
{\em American Statistician}, 49,{\bf 1}, 48-53.\vspace*{-0.25cm}
%
%
\item McEwan P. J. (2000). The Potential Impact of Large-Scale Voucher
Programs. {\em Review of Educational Research}, {\bf 70},
103-149.\vspace*{-0.25cm}
\item Morgan S. L. and Winship C. (2007). Counterfactual and Causal
inference. Methods and Principles for Social Research. Cambridge
University Press.\vspace*{-0.25cm}
\item National Assessment of Educational Progress (2005). {\em The Nation's
Report Card Student Achievement in Private Schools}. United States
Department of Education, Institute of Education Sciences NCES
2006-459.\vspace*{-0.25cm}
\item National Assessment of Educational Progress (2006).  {\em Comparing
Private Schools and Public Schools Using Hierarchical Linear
Modeling}. United States Department of Education, Institute of
Education Sciences NCES 2006-461.\vspace*{-0.25cm}
\item OECD (2007). Executive Summary PISA (2006): Science Competencies for Tomorrow's World.
OECD report. Available at\newline
\texttt{http://www.oecd.org/dataoecd/15/13/39725224.pdf}.\vspace*{-0.25cm}
\item Rasch, G. (1961). On general laws and the meaning of measurement in
psychology.  {\em Proceedings of the IV Berkeley Symposium on
Mathematical Statistics and Probability}, {\bf 4},
321-333.\vspace*{-0.25cm}
\item Raudenbush, S.W. and Bryk A. (2002). {\em Hierarchical Linear Models:
Applications and Data Analysis Methods}.(Second edition), Newbury
Park, CA: Sage Publications. \vspace*{-0.25cm}
%
%
%
\item Rubin, D.B., Stuart, E.A. and Zanutto, E.L. (2004). A potential
outcomes view of value-added assessment in education.  {\em Journal
of Educational and Behavioral Statistics}, {\bf 29}, Value-Added
Assessment Special Issue, pp. 103-116.\vspace*{-0.25cm}
\item Schneider B., Carnoy M., Kilpatrick J., Schmidt W. H. and Shavelson
R. J. (2007). Estimating Causal Effects Using Experimental and
Observational Designs. {\em A Think Tank White Paper}, Washington
Dc: American Educational Research Association.\vspace*{-0.25cm}
\item Schwarz, G. (1978). Estimating the dimension of a model,
{\it Annals of Statistics}, {\bf 6}, 461-464.
\item Snijders, T.A.B. and Bosker, R. J. (1999). {\em Multilevl
Analysis}. London: Sage Publications.
\item Stuart E. A. (2007). Estimating Causal Effects Using School-Level
Data Sets. {\em Educational Researcher}, {\bf 36},
187-198.\vspace*{-0.25cm}
\item Vermunt, J. K. Langeheine, R., Böckenholt, U. (1999).
Discrete-time discrete state latent Markov models with time-constant
and time-varying covariates. {\em Journal of Educational and
Behavioral Statistics}, {\bf 24}, 178-205.\vspace*{-0.25cm}
\item Wiggins, L. M. (1973). {\em Panel Analysis: Latent
Probability Models for Attitudes and Behavior Processes}. Amsterdam:
Elsevier
\end{description}}

\newpage

\vspace*{0.25cm}

\begin{table}[h]\centering{
{\small
\begin{tabular}{lccc|ccc}\\ \hline
 &             & \multicolumn2c{$public$} & \multicolumn2c{$paritarie$}   \\ \hline
 &             & $\%$         & cum  $\%$    & $\%$  & cum$\%$\\
\it{number of students}       &         &       &        \\
 & (0-200)     & 49.02        & 49.02   & 82.59  & 82.59   \\
 & (200-350)   &  31.30       & 80.31   & 17.41 & 100.00  \\
 &  (350-700)  &  15.55       & 95.87   & 00.00 & 100.00  \\
 & (700-1050)  & 4.13         & 100.00  & 00.00 & 100.00   \\
\it{number of teachers}       &         &       &    \\
 & (0-20)     & 20.90         & 20.90   & 53.86 & 53.86   \\
 & (20-40)    & 32.52         & 53.42   & 46.14 & 100.00 \\
 &  (40-70)   & 30.63         & 84.05   & 00.00 & 100.00  \\
 & (70-105)   & 15.95          & 100.00  & 00.00 & 100.00   \\
 \it{students-teachers ratio}       &         &       &    \\
 & (1-6)     & 6.50          & 6.50     & 11.13  & 11.13   \\
 & (6-8)     & 22.05         & 28.54    & 25.12 & 36.25 \\
 &  (8-12)   & 64.17         & 92.72    & 17.20 & 53.45  \\
 & (12-20)   & 7.28          & 100.00   & 45.55  & 100.00   \\
  \\ \hline
\end{tabular}}}
\caption{\small\em  Frequency distributions of the number of
students and teachers and their ratio at the school level for the
public and paritarie middle schools in the selected
areas.}\label{Table1}
\end{table}
\begin{table}[h]\centering{
{\small
\begin{tabular}{lccc|ccc}\\ \hline
 &             &  \multicolumn2c{$public$}         &  \multicolumn2c{$paritarie$}   \\ \hline
 &             & $\%$         & cum  $\%$    & $\%$  & cum$\%$ \\
\it{number of students}       &         &       &        \\
 & (0-200)     &  15.38  &   15.38  & 85.71        & 85.71 \\
 & (200-350)   &  30.77  &  46.15   & 14.29       & 100.00 \\
 &  (350-700)  & 30.77   & 76.92    & 0.000       & 100.00 \\
 & (700-1050)  & 23.08   & 100.00   & 0.000        & 100.00 \\
\it{number of teachers}       &         &       &    \\
 & (0-20)      & 15.38   & 15.38    & 57.14        & 57.14    \\
 & (20-40)     & 30.77   & 46.15    & 42.86         & 42.86  \\
 &  (40-70)    & 38.46   &84.62     & 0.000         & 0.000  \\
 & (70-105)    & 15.38   & 100.00   & 0.000         & 0.000 \\
 \it{students-teachers ratio}       &         &       &    \\
 & (1-6)       & 0.000   & 0.000    & 14.29           & 14.29   \\
 & (6-8)       & 30.77   & 30.77    &  28.57         &  42.86  \\
 &  (8-12)     & 61.54   &  92.31   & 14.29          & 57.14 \\
 & (12-20)     & 7.69    & 100      &  42.86          & 100.00  \\
\it{years since school opened}       &         &       &    \\
 & $\leq 17.5$         & 69.20         & 69.20   & 71.40   & 71.40  \\
 & $> 17.5$        &  30.80         &  100.00    & 28.60   & 100.00   \\
 \\ \hline
\end{tabular}}}
\caption{\small\em Frequency distributions of the number of
students, teachers and their ratio and the years the school has been
in operation for the public and paritarie middle schools included in
the observed sample.}\label{Table2}
\end{table}
\begin{table}[h]\centering{
{\small
\begin{tabular}{llcc|ccc}\\ \hline
 &                   &  $public$   &  $paritarie$     &      \\ \hline
  &             & $\%$         &  $\%$    & total  \\
\it{Father's education}       &         &       &        \\
 & no response    &  7.34     &  14.29  & 8.75        \\
 & primary school          & 3.92  &  1.59   & 3.45      \\
 & middle school  & 25.96   & 9.13    & 22.25       \\
 & high school  & 38.93   & 35.71   & 38.28     \\
 & college degree or higher  & 23.84   & 39.29   & 26.97       \\
\it{Mother's education}       &         &       &    \\
 & no response     &6.64   & 12.30   & 57.14         \\
 &  primary school         & 3.62   & 0.79    & 3.05         \\
 &  middle school        & 25.75   &9.52     & 22.47       \\
 &  high school      & 44.06   & 40.87  & 43.42          \\
 & college degree or higher  & 19.92   & 36.51   & 23.27         \\
 \\ \hline
\end{tabular}}}
\caption{\small\em  Frequency distribution of parental education of
public and paritarie middle school students included in the
sample.}\label{Table3}
\end{table}

\begin{table}[h]\centering
{\small
\begin{tabular}{cccccccc}
\hline    $k_1$ &  $k_2$   & $\hat{\ell}_{k_1,k_2}$    & $BIC_{k_1,k_2}$   & $np$
\\\hline
1   &   1   &   -76565.77   &   153737.40   &   85  \\
1   &   2   &   -71340.91   &   143415.97   &   103 \\
1   &   3   &   -70275.94   &   141357.31   &   113 \\
1   &   4   &   -69878.97   &   140663.16   &   127 \\
1   &   5   &   -69703.14   &   140439.79   &   145 \\
1   &   6   &   -69703.14   &   140439.79   &   145 \\
1   &   7   &   -69561.12   &   140497.88   &   193 \\
2   &   2   &   -71266.31   &   143316.67   &   110 \\
2   &   3   &   -70187.30   &   141229.93   &   120 \\
2   &   4   &   -69771.10   &   140497.30   &   134 \\
2   &   5   &   -69579.94   &   140243.28   &   152 \\
2   &   6   &   -69490.77   &   140221.77   &   174 \\
2   &   7   &   -69425.48   &   140276.51   &   200 \\
3   &   2   &   -70187.30   &   141229.93   &   120 \\
3   &   3   &   -70128.67   &   141162.56   &   127 \\
3   &   4   &   -69707.65   &   140420.31   &   141 \\
3   &   5   &   -69515.20   &   140163.70   &   159 \\
3   &   6   &   -69410.43   &   140110.98   &   181 \\
3   &   7   &   -69349.18   &   140173.78   &   207 \\
4   &   2   &   -71204.06   &   143291.96   &   124 \\
4   &   3   &   -70093.64   &   141142.39   &   134 \\
4   &   4   &   -69664.37   &   140383.63   &   148 \\
4   &   5   &   -69482.37   &   140147.94   &   166 \\
{\bf 4}    &      {\bf 6}    &      {\bf -69374.80}    &    {\bf 140089.61}    &  {\bf 188} \\
4   &   7   &   -69315.71   &   140156.75   &   214 \\
5   &   2   &   -71184.70   &   143303.13   &   131 \\
5   &   3   &   -70073.61   &   141152.23   &   141 \\
5   &   4   &   -69631.02   &   140366.83   &   155 \\
5   &   5   &   -69457.00   &   140147.09   &   173 \\
5   &   6   &   -69356.58   &   140103.06   &   195 \\
5   &   7   &   -69297.09   &   140169.39   &   221 \\
\\\hline
\end{tabular}}
\caption{\small\em  For any number of latent states at cluster level
($k_1$) and at individual level ($k_2$), $\hat{\ell}_{k_1,k_2}$ is
the corresponding maximum log-likelihood, $BIC_{k_1,k_2}$ is the
corresponding BIC index and $np$ the number of parameters. Figures
in boldface correspond to the model with the smallest value of the
BIC index.}\label{table4}
\end{table}
\begin{table}[h]\centering
{\small
\begin{tabular}{llcccc}
\hline  \hline
$v$ & $\th_v$ \\
\hline
   1     & $0.000$  \\
   2    & $0.866$  \\
   3    & $1.800$   \\
   4     & $2.698$ \\
   5    & $3.623$   \\
   6    & $4.825$   \\
   \hline
\end{tabular}}
\caption{\small\em Estimated math ability levels across all groups
 of students.} \label{table5}
\end{table}

\begin{table}[h]\centering
\begin{tabular}{llllll}
\hline    \hline & & estimates & s.e. & $p-value$ \\ \hline
& $\ga_{0C}$ & -42.030 & - & -\\
& $\ga_{1B1}$ & 28.815 & 17.481 & 0.099  \\
& $\ga_{1B2}$ & 1.263 & 0.417 & 0.002 \\
& $\ga_{1B3}$ & -1.283 & 1.007 & 0.203 \\  \hline
& $\ga_{0C}$ & -31.359 & - & -  \\
& $\ga_{1C1}$ & 28.389 & 10.144 & 0.005 \\
& $\ga_{1C2}$ & 0.336 & 0.231 & 0.145\\
& $\ga_{1C3}$ & -0.574 & 0.973 & 0.555\\  \hline
& $\ga_{0D}$ & 2.851 & - & - \\
& $\ga_{1D1}$ & -0.039 & 0.967 &0.968 \\
& $\ga_{1D2}$ & -0.620 & 0.221  & 0.005  \\
& $\ga_{1D3}$ & 2.754 & 1.154  &0.017 \\ \hline
\end{tabular}
\caption{\small\em  Estimated intercepts and  regression
coefficients of the logit defined on the group level latent variable
for the covariate type of school labeled with 1 (public or
paritaria),  the ratio between students and teachers labeled with 2
and for the dummy variable
years since school opened labeled with 3 for any number of group
level latent class $u = A, B,C, D$.}\label{table6}
\end{table}
\begin{table}[h]\centering
\begin{tabular}{ccccc}
\hline  & $A$  & $B$  & $C$  & $D$\\\hline
 {\em public} &0.325  & 0.233 & 0.376 & 0.066 \\
{\em paritarie}   & 0.786  &  0.000 & 0.000  & 0.214\\\hline
{\em $\geq 17.5$ years} &0.425  & 0.189 & 0.314 & 0.072 \\
{\em $< 17.5$ years}    & 0.363  &  0.196 & 0.289  & 0.152\\ \hline
{\em $< 8$}         &0.541 & 0.090 & 0.205 & 0.164 \\
{\em $\geq 8 $}     & 0.337  &  0.245 & 0.363  & 0.054\\
\hline
\end{tabular}
\caption{\small\em Average class probabilities among type of school,
years since school opened and ratio between students and teachers of
belonging to each latent class of the group latent variable $U$.
}\label{table7}
\end{table}

\begin{table}[h]\centering
\begin{tabular}{llllll}
\hline \hline &   & estimates & s.e & $p$-value \\ \hline
& $\de_{0B}$ & - 0.261  & - &- \\
& $ \de_{0C}$  & 1.140  & - &-  \\
& $\de_{0D} $  & 0.013  &-  & - \\ \hline
&  $ \de_{2,mF}$   &  1.058 & 3.273  & 0.747 \\
&  $\de_{2,F}$  & 0.403 & 0.143  & 0.005 \\
& $ \de_{2,mM}$   &  0.139 & 0.054 & 0.011 \\
&   $\de_{2,M}$ & 0.292 & 0.067 & 0.000\\ \hline
\end{tabular}
\caption{\small\em  Estimated intercepts for each cluster level of
the global logit defined on the initial probabilities. Estimated
regression parameters common for each ability level  for the
covariate father's education  labeled with  F and  mother's
education labeled with M  and for the dummy variable for missing
responses labeled with mF and mM respectively.}\label{table8}
\end{table}

\begin{table}[h]\centering
\begin{tabular}{llllll}
\hline \hline  &   & estimates & s.e &$p$-value \\ \hline
& $\eta_{0B}^2$   & -2.774  &- &-\\
& $\eta_{0C}^2$   &-0.317 &- &- \\
& $\eta_{0D}^2$   & 4.515 &- &- \\  \hline
& $\eta_{0B}^3$   & 3.893 &- & -\\
& $\eta_{0C}^3$   & 1.673 &- &-\\
& $\eta_{0D}^3$   & -0.434 &- &-\\ \hline
& $\eta_{2,mF}^2$   & -0.675 &0.349 & 0.053\\
&  $\eta_{2,F}^2$  & 0.281 &0.141 & 0.046\\
& $\eta_{2,mM}^2$  & 0.581 &0.670 & 0.386\\
& $\eta_{2,M}^2$   & 0.007 &0.164 & 0.965 \\ \hline
& $\eta_{2,mF}^3$  & 0.867 & 0.414 & 0.036 \\
& $\eta_{2,F}^3$   & 0.320 &0.125 & 0.011\\
& $\eta_{2,mM}^3$  & 0.765 & 0.135 & 0.015\\
&  $\eta_{2,M}^3$   & 0.143 &0.172 & 0.406\\ \hline \hline
\end{tabular}
\caption{\small\em Estimated parameters affecting the transition
probabilities from Grade 6 to 7 (2) and Grade 7 to 8 (3)   of the
latent Markov process. Estimated effects of the clusters and
estimated regression coefficients of the variables father's and
mother's education labeled with F and M respectively and of the
dummy variable for the missing responses labeled with mF and mM
respectively}\label{table9}
\end{table}

\begin{table}[h]\centering
{\small\begin{tabular}{ccccccc|c}
\hline  & \multicolumn5c{$Grade$ $7$}\\
\cline{2-6}  $Grade$ $6$ & $1$ & $2$ & $3$ & $4$ & $5$  & $6$  & $total$\\
\hline
1   & 0.11       & {\bf\em 0.67}  &$0.22$        & $0.00$          & $0.00$      & 0.02       &  1.00    \\
2   & $0.05$     & 0.36   &{\bf\em 0.37}         & $0.16$          & $0.06$      & 0.00       & 1.00    \\
3   & 0.02    &$0.21$             & {\bf\em 0.44}    & $0.23$      & $0.07$      & 0.02        & 1.00  \\
4   & 0.00    &$0.09$             &0.27           &  {\bf\em 0.35} & $0.24$      & 0.04      &  1.00 \\
5   & $0.00$  & 0.02              & 0.19         &  0.31  &  {\bf\em 0.35}        & 0.14     & 1.00  \\
6  & $0.01$   & {0.02}            &0.04          &   0.15          &
{\bf\em 0.40}    & 0.35   & 1.00\\\hline
    \hline
\end{tabular}}

{\small\begin{tabular}{ccccccc|c}
\hline  & \multicolumn5c{$Grade$ $8$}\\
\cline{2-6}  $Grade$ $7$ & $1$ & $2$ & $3$ & $4$ & $5$ &6 &  $total$\\
\hline
1   & 0.00    & {\bf\em 0.53}    &$0.18$        & $0.06$       & $0.06$   & 0.18       &  1.00    \\
2   & 0.03     &{\bf\em 0.38}    & 0.35         & $0.21$       & $0.02$   & 0.01       & 1.00   \\
3   & 0.01    & 0.22             & 0.30    & {\bf\em 0.32}     & $0.13$   & 0.03       & 1.00  \\
4   & 0.00    &0.07              & 0.24       & {\bf\em 0.42}  & $0.23$   & 0.05       & 1.00  \\
5   & 0.00     & 0.01            & 0.04        & 0.23          &{\bf\em 0.30 }  & 0.14     & 1.00 \\
6  & 0.00     & {0.03}           &0.17       &   0.35
&{\bf\em  0.44}    &0.00         & 1.00\\\hline
    \hline
\end{tabular}}
\caption{\small\em Empirical  transition probabilities from Grade 6
to Grade 7 and  from Grade 7 to Grade 8 for students attending
public school. Figures  in italic and boldface correspond to the
largest probability in any row.} \label{Table10}
\end{table}

\begin{table}[h]\centering
{\small\begin{tabular}{ccccccc|c}
\hline  & \multicolumn5c{$Grade$ $7$}\\
\cline{2-6}  $Grade$ $6$ & $1$ & $2$ & $3$ & $4$ & $5$  & $6$  & $total$\\
\hline
1   & 0.00     & 0.33  & {\bf\em 0.67}   & $0.00$              & $0.00$     & 0.00        &  1.00    \\
2   & 0.03     & 0.26  &{\bf\em  0.28}            & {\bf\em 0.28}       & $0.13$     & 0.03        &  1.00   \\
3   & 0.02     & 0.11  & {\bf\em 0.41}   & 0.27                & $0.12$     & 0.08        &  1.00 \\
4   & 0.00     & 0.02  &0.26             & {\bf\em 0.38}       & $0.24$     &  0.11       &   1.00  \\
5   & $0.00$   & 0.02  & 0.30            & {\bf\em 0.45 }                &  0.23  & 0.00      &  1.00  \\
6  & $0.00$    & 0.00  &0.00         &  0.00                   &0.43
&{\bf\em 0.57}     &  1.00\\\hline
    \hline
\end{tabular}}

{\small\begin{tabular}{ccccccc|c}
\hline  & \multicolumn5c{$Grade$ $8$}\\
\cline{2-6}  $Grade$ $7$ & $1$ & $2$ & $3$ & $4$ & $5$ &6 &  $total$\\
\hline
1   & 0.50     & {\bf\em 0.50}  & $0.00$        & $0.00$         & 0.00          & 0.00       & 1.00    \\
2   & 0.10     &{\bf\em 0.60}   & 0.30          & $0.00$        & $0.00$         & 0.00       & 1.00    \\
3   & 0.00    &0.25             & {\bf\em 0.45} & $0.28$        & $0.03$         & 0.00       & 1.00  \\
4   & 0.00    &0.03             &0.27            & {\bf\em 0.51} & $0.19$        &  0.00      & 1.00 \\
5   & 0.00     & 0.02           &0.05           & 0.26            & {\bf\em 0.53} & 0.14      & 1.00  \\
6  & 0.00     & 0.00           &0.00            & 0.07           &
0.17       &  {\bf\em 0.28}   & 1.00\\\hline
    \hline
\end{tabular}}
\caption{\small\em Empirical  transition probabilities from Grade 6
to Grade 7 and from Grade 7 to Grade 8 for students attending
paritaria school. Figures  in italic and boldface correspond to the
largest probability in any row.} \label{Table11}
\end{table}

\end{document}